\documentclass[prd,preprint,superscriptaddress]{revtex4}
\usepackage{revsymb}
\usepackage{amsmath}
\usepackage{graphicx}
\usepackage{bm}
\usepackage{amsmath}
\newcommand{\be}{\begin{equation}}
\newcommand{\ee}{\end{equation}}
\newcommand{\ben}{\begin{eqnarray}}
\newcommand{\een}{\end{eqnarray}}

\begin{document}
\title{Interacting holographic dark energy}
\date{\today}
\author{Winfried Zimdahl\footnote{E-mail: zimdahl@thp.uni-koeln.de}}
\affiliation{
Departamento de F\'{\i}sica, Universidade Federal do Esp\'{\i}rito
Santo, CEP29060-900 Vit\'oria, Esp\'{\i}rito Santo, Brazil }
\author{Diego Pav\'{o}n\footnote{E-mail address: diego.pavon@uab.es}}
\affiliation{Departamento de F\'{\i}sica, Facultad de Ciencias,
Universidad Aut\'{o}noma de Barcelona, 08193 Bellaterra
(Barcelona), Spain}

\begin{abstract}
We demonstrate that a transition from decelerated to accelerated
cosmic expansion arises as a pure interaction phenomenon if
pressureless dark matter is coupled to holographic dark energy
whose infrared cutoff scale is set by the Hubble length. In a
spatially flat universe the ratio of the energy densities of both
components remains constant through this transition, while it is
subject to slow variations for non-zero spatial curvature. The
coincidence problem is dynamized and reformulated in terms of the
interaction rate. An early matter era is recovered since for
negligible interaction at high redshifts the dark energy itself
behaves as matter.  A simple model for this dynamics is shown to
fit the SN Ia data. The constant background energy density ratio
simplifies the perturbation analysis which is characterized by
non-adiabatic features.
\end{abstract}
\maketitle

\section{Introduction}
Nowadays an overwhelming, direct and indirect, observational
evidence supports the idea  that the Universe is currently
undergoing a phase of accelerated expansion. The most recent and
precise confirmation of this rather unexpected feature (see,
however \cite{Turner97,Priester})  was provided by the 3rd year
data of the WMAP mission \cite{WMAP3}. Likewise, data from
high--redshift supernovae type Ia \cite{SNIa}, the cosmic
microwave background radiation \cite{cmb}, the large scale
structure \cite{lss}, the integrated Sachs--Wolfe effect
\cite{isw}, and weak lensing \cite{weakl}, endorse it to the point
that the discussion has now shifted to when the acceleration began
and which might be the agent behind it. For recent reviews see
\cite{recentrev}.

According to our understanding on the basis of Einstein's gravity,
more than $70$\% of the cosmic substratum, dubbed  dark energy,
must be endowed with a high negative pressure. Less than $30$\% is
found to be pressureless matter. Most of the latter is in the form
of cold dark matter; only about $5$\%  corresponds to ``normal"
baryonic matter. Currently, we neither know what the dark matter
is made of nor do we understand the nature of dark energy.

Lacking a fundamental theory, most investigations in the field are
phenomenological and rely on the assumption that these two unknown
substances evolve independently, i.e., their energies are assumed
to obey separate conservation laws. In particular, this implies
that the dark matter energy density varies as $a^{-3}$, where $a$
is the scale factor of the Robertson-Walker metric. The behavior
of the density of the dark energy is then entirely governed by its
equation of state, the determination of which is a major subject
of current observational cosmology. One should realize, however,
that any coupling in the dark sector will change this situation. A
coupling will modify the evolution history of the Universe. As one
of the consequences, the energy density of the (interacting) dark
matter will no longer evolve as $a^{-3}$. Ignoring a potentially
existing interaction between dark matter and dark energy from the
start, may result in a misled interpretation of the data
concerning the dark energy equation of state. Das {\em et al.}
\cite{Das} and Amendola {\em et al.} \cite{Luca} have shown that a
measured phantom equation of state may be mimicked by an
interaction, while the bare equation of state may well be of a
non-phantom type. Further, models showing interaction fare well
when contrasted with data from the cosmic microwave background
\cite{german1} and matter distribution at large scales
\cite{german2}. It can therefore be argued that the possibility of
dark energy to be in interaction with dark matter must be taken
seriously.

On the other hand, there exist limits for the strength of this
interaction for various configurations \cite{limits}. It is
characteristic for these approaches that they admit a
non-interacting limit. If the interaction is switched off, they
continue to represent models of a mixture of dark matter and dark
energy. The best known examples are models with a decaying
cosmological ``constant": if the decay is switched off, they
reduce to the $\Lambda$CDM model. An interaction here leads to
(possibly important) corrections of the non-interacting
configuration. E.g., for a given (not necessarily constant)
negative equation of state parameter of the dark energy, the
interaction manifests itself in third order in the redshift in the
luminosity distance of supernovae type Ia \cite{WDGRG}. It does
not, however, (directly) influence the leading orders.

The qualitatively new aspect of the present approach is the
circumstance that an interaction is crucial already in leading
order. In this context the accelerated expansion itself is a
consequence of the interaction. The coupling does not just lead to
higher order corrections of a non-interacting reference model
which by itself provides an accelerated expansion of the Universe.
The non-interacting limit of the present approach reproduces an
Einstein-de Sitter universe, i.e., there is no accelerated
expansion if the interaction is switched off. This non-interacting
limit  is supposed to characterize our Universe at high redshifts.

The data also suggest that the present Universe is nearly
spatially flat \cite{WMAP3}, \cite{seljak}.  Many studies neglect
the spatial curvature term and focus solely on the spatially flat
case, thereby taking for granted that the spatial curvature term
necessarily leads to trifling corrections. Several of the not so
many works that retain that term, for the sake of generality, even
seem to justify that it is of minor importance. On the other hand,
thanks to the increasing observational precision also higher order
corrections may become within reach in the not too distant future.
The authors of \cite{Caldwell} have demonstrated that the spatial
curvature enters the luminosity distance of SNIa supernovae in
third order in redshift (see also \cite{WangGongSU}). It is also
known that there are degeneracies between the curvature and the
dark energy equation of state in the corresponding parameter space
\cite{Degeneracies}. The possible relevance of spatial curvature
for the lowest multipoles in the cosmic microwave background
radiation was discussed in \cite{Abdalla}. These examples indicate
that apart from general theoretical grounds, the observational
situation will require the inclusion of spatial curvature, even if
its contribution is small, at some level of precision. Further, as
demonstrated by Ichikawa {\it et al.} \cite{Ichikawa}, depending
on the parametrization of the equation of state parameter the
present value of the spatial curvature can be as large as $0.2$.
Therefore, the additional dynamics provided by the curvature,
should certainly not be dismissed too quickly since it contains
information which is needed to further restrict the cosmological
parameter space. In this connection, Clarkson {\it et al.}
\cite{Bassett} have shown that by excluding the spatial curvature
when interpreting the empirical data one can incur  in  gross
mistakes when reconstructing the equation of state of dark energy.
We mention that an effective spatial curvature term also appears
as the result of an averaging procedure within the ``macrosopic
gravity" approach \cite{Zala} (for a further discussion of the
effects of averaging on cosmological observations see, e.g.,
\cite{alan}).

In this paper we consider pressureless dark matter in interaction
with an unknown component which is supposed to describe dark
energy. We neither specify the dark energy equation of state nor
the interaction rate from the beginning. These two (generally time
dependent) parameters will influence the ratio of the energy
densities of dark matter and dark energy. The behavior of this
ratio is crucial for  the ``conventional" form of the
``coincidence problem", namely: ``why are the matter and dark
energy densities of precisely the same order today". In principle,
matter and dark energy redshift at different rates. We show, that
there exists a preferred class of dark energy models for which the
dynamics of the energy density ratio is entirely determined by the
spatial curvature. For vanishing curvature the energy density
ratio remains constant. These models are singled out by a
dependence $\rho_{X} \propto H^{2}$, where $\rho_{X}$ is the dark
energy density and $H = \dot{a}/a$ is the Hubble parameter.
Exactly this dependence is characteristic for a certain type of
dark energy models, inspired by the holographic principle
\cite{hooft,leonard}.

Holographic dark energy models must specify an infrared cutoff
length scale \cite{cohen}. The choice of this scale is presently a
matter of debate. The most obvious choice, the Hubble length,
seemed to be incompatible with an accelerated expansion of the
Universe \cite{Hsu} (see, however, \cite{raul1,guberina}). This is
why, starting with the work of Li \cite{mli1}, many researchers
have adopted the future event horizon as the cutoff scale as this
choice allows for a sufficiently negative equation of state
parameter and hence an accelerated expansion -see, e.g.
\cite{feventh}.

In a previous paper \cite{DW}, we showed that a cutoff set by the
Hubble length may well be compatible with an accelerated expansion
provided  that the dark energy and dark matter do not evolve
separately but interact, also non--gravitationally, with each
other. In this setting, a negative equation of state parameter
that gives rise to accelerated expansion arises as a direct
consequence of the interaction. Here we put this feature in a
broader context and  demonstrate that a constant or slowly varying
(as the consequence of a non-vanishing spatial curvature) energy
density ratio is compatible with a transition from decelerated to
accelerated expansion under the condition of a growing interaction
parameter. We show that in holographic dark energy models with a
Hubble length cutoff a transition from decelerated to accelerated
expansion is realized as a pure interaction effect. At high
redshifts and for negligible interaction the dark energy equation
of state approaches the equation of state for matter, such that a
preceding matter era is naturally recovered.  In this context the
coincidence problem can be rephrased as follows: ``why is the
interaction rate between dark matter and dark energy of the order
of the Hubble rate precisely at the present epoch?" This
reformulation allows us to address the coincidence problem as part
of the interaction dynamics in the dark sector.

Different interaction rates will imply a different perturbation
dynamics. We shall demonstrate that non-adiabatic features appear
as a characteristic signature for a large class of interacting
models. These effects may be used to discriminate between
different models of the cosmic medium which share the same
background dynamics.

The outline of this paper is as follows. Section \ref{general}
provides the general formalism for an interacting two-component
fluid, where one of the fluids is pressureless. Then it focuses on
the case for which the ratio of the energy densities of both
fluids is constant or slowly varying. This will single out models
for which the energy density of the second component is
proportional to the square of the Hubble parameter. A realization
of this dependence is provided by certain holographic dark energy
models. The basic properties of these models are recalled in
Section \ref{holography}. Section \ref{interaction} discusses in
detail matter in interaction with holographic dark energy. It
provides the conditions under which an interaction driven
transition from decelerated to accelerated expansion is feasible.
It also comments on the possibility of a weak time dependence of
the saturation parameter of the holographic bound, which is
usually assumed constant. Section \ref{model} introduces a simple
interacting model and compares it with the $\Lambda$CDM model. The
perturbation dynamics for a fluctuating interaction rate is
discussed in \ref{perturbations} with special emphasis on
non-adiabatic aspects on large perturbation scales. Finally,
section \ref{discussion} summarizes our findings.

\section{Interacting cosmological fluids}
\label{general}

The field equations for a spatially homogeneous and isotropic
universe are the Friedmann  equation
\begin{equation}
3\,H^{2} = 8\,\pi\,G\,\rho - 3\,\frac{k}{a^{2}} \, , \qquad \quad
 (k =+1, 0, -1)\, ,
 \label{friedmann}
\end{equation}
and
\begin{equation}
\dot{H}\, = - 4\,\pi\,G\,\left(\rho + p\right)  + \frac{k}{a^{2}}
\ .\label{dotH}
\end{equation}
A combination of both equations  yields
\begin{equation}
2\frac{\dot{H}}{H}\, = - 3H\left(1 + \frac{p}{\rho}\right) -
\frac{k}{a^{2}H}\left(1 + 3\frac{p}{\rho}\right) \
,\label{2dotH/H}
\end{equation}
a relation that will be useful later on. The deceleration
parameter, $q = -\ddot{a}/(aH^{2})$, can be written as
\begin{equation}
q =  \frac{1}{2}\left(1 + \frac{k}{a^{2}H^{2}}\right)\left(1 +
3\frac{p}{\rho}\right)
 \ .\label{q2}
\end{equation}
\\
The total energy density $\rho$ is supposed to split into $\rho =
\rho_{M} + \rho_{X}$, where $\rho_{M}$ is the energy density of
pressureless dark matter. Under this assumption the total pressure
equals the dark energy pressure, $p = p_{X}$. We further assume
that both components do not conserve separately but interact with
each other in such a manner that the balance equations take the
form
\begin{equation}
\dot{\rho}_{M} + 3H \rho_{M} = Q\, \,  \quad \mbox{and} \quad
\dot{\rho}_{X} + 3H (1+w)\rho_{X} = - Q \, , \label{dotrhom}
\end{equation}
where $w \equiv p_{X}/\rho_{X}$ is the equation of state parameter
of the dark energy, and the function $Q > 0$ measures the strength
of the interaction.  Models featuring an interaction matter--dark
energy were introduced by Wetterich \cite{wetterich} (see also
\cite{andrew}) and first used alongside the holographic dark
energy by Horvat \cite{raul1}. Nowadays there is a growing body of
literature on the subject -see, e.g. \cite{interacting} and
references therein. Although the assumption of a coupling between
both components implies the introduction of an additional
phenomenological function $Q$, a description that admits
interactions is certainly more general than otherwise. Further,
there is no known symmetry that would suppress such interaction
and arguments in favor of interacting models have been put forward
recently \cite{farrar}.

The quantity of interest for analyzing the coincidence problem is
the ratio $ r \equiv \rho_{M}/\rho_{X}$, which, upon using Eqs.
(\ref{dotrhom}), can be written  as
\\
\begin{equation}
\dot{r} = \left(1 + r\right)\left[3\,H\,w\,\frac{r}{1 + r} +
\Gamma \right]\ .\label{rdotalt2a}
\end{equation}
\\
Here we have introduced the quantity $\Gamma \equiv Q/\rho_{X}$
which characterizes the rate by which $\rho_{X}$ changes as a
result of the interaction. On the other hand, combining the
balance equation for $\rho_{X}$ in (\ref{dotrhom}) with
Eq.~(\ref{2dotH/H}), we obtain
\\
\begin{equation}
\frac{\dot{\rho}_{X}}{\rho_{X}} - 2\,\frac{\dot{H}}{H} = -
\left[3H\,w\,\frac{r}{1 + r} + \Gamma\right] +
\frac{k}{a^{2}\,H}\left(1 + 3\frac{w}{1 + r}\right)\
.\label{dotX-dotH}
\end{equation}
\\
Comparing now Eqs.~(\ref{dotX-dotH}) and (\ref{rdotalt2a}), it
follows that the dynamics of the aforesaid ratio is governed by
\\
\begin{equation}
\dot{r} = - \left(1 +
r\right)\left[\frac{\dot{\rho}_{X}}{\rho_{X}} -
2\,\frac{\dot{H}}{H} - \frac{k}{a^{2}\,H}\left(1 +
3\,\frac{w}{1+r}\right)\right]\, . \label{rdotalt2}
\end{equation}
\\
It is this formula which deserves attention with regard to the
coincidence problem. Clearly, the lower $|\dot{r}|$, the less
acute this problem seems to be. Inspection of Eq.~(\ref{rdotalt2})
shows that the case $\rho_{X} \propto H^{2}$ is singled out for it
leads to
\\
\begin{equation}
\qquad\qquad \dot{r} =  \left(1 + r\right)
\,\frac{k}{a^{2}\,H}\left(1 + 3\,\frac{w}{1+r}\right) \, .
\label{rdotalt3}
\end{equation}
\\
At this point, it is expedient to introduce the dimensionless
quantities
\\
\begin{equation}
\Omega_{M} = \frac{8\pi \, G \,\rho_{M}}{3\, H^{2}}, \qquad
\Omega_{X} = \frac{8\pi\, G \, \rho_{X}}{3\, H^{2}}, \qquad
\Omega_{k} = - \frac{k}{a^{2}\, H^{2}}. \label{defOmega}
\end{equation}
Thus, Friedmann's equation (\ref{friedmann}) can be cast as
\begin{equation}
\Omega_{M} + \Omega_{X} + \Omega_{k} = 1  \, .
\label{sumOmega}
\end{equation}
By using it together with Eq.~(\ref{q2}), the expression
(\ref{rdotalt3}) takes the form
\begin{equation}
\dot{r} = - 2\, H\,\frac{\Omega_{k}}{\Omega_{X}}\,q\
.\label{rdotalt4}
\end{equation}
Via the curvature the evolution of $r$ is directly linked to the
deceleration parameter. For $k= +1$ and $q>0$ (decelerated
expansion) $r$ augments, and for $q<0$ (accelerated expansion) $r$
diminishes. For $k=-1$ the behavior is just the opposite. In any
case, only small variations of $r$ are possible nowadays if
$1/(a^{2}\,H^{2}) \ll 1$ at present \cite{WMAP3,seljak}. While a
slow time dependence of the energy density ratio is desirable from
the point of view of the coincidence problem, it remains to be
clarified whether dark energy models with $\rho_{X} \propto H^{2}$
are able to account for a present phase of accelerated expansion
as well as for a transition to the latter from an earlier matter
dominated phase.

A survey of the, by now, rather ample body of literature on dark
energy candidates reveals that models of this type have indeed
been introduced previously, namely, in the context of ideas which
are rooted in the holographic principle. In the following section
we briefly recall, how a dependence $\rho_{X} \propto H^{2}$
emerges in these holographic dark energy models.

\section{Holographic dark energy}
\label{holography}

The basic ideas of the holographic principle were introduced by `t
Hooft \cite{hooft} and Susskind \cite{leonard}. The development of
interest for our purpose was put forward by Cohen {\it et al.}
\cite{cohen}, followed by Hsu \cite{Hsu} and Li \cite{mli1}, who
considered specific holographic dark energy models.

The essential point in establishing the holographic idea is the
way of counting the degrees of freedom of a physical system.
Consider a three-dimensional lattice of spin-like degrees of
freedom and assume that the distance between every two neighboring
sites is some small length $\ell$ which is of the order of the
Planck length, $\ell_{Pl}$. Each spin can be in one of two states.
In a region of volume $L^{3}$ the number of quantum states $N$
will be $N = 2^{n}$, with $n= (L/\ell)^{3}$ the number of sites in
the volume, whence the entropy, given by the logarithm of $N$,
will be $S \propto (L/\ell)^{3} \ln 2$. Identifying (in Planck
units) $\ell^{-1}$ with the ultraviolet cutoff $\Lambda$ (the
corresponding energy is $L^{3}\Lambda^{4}$), the maximum entropy
varies as $S \sim L^{3} \, \Lambda^{3}$, i.e., proportional to the
volume of the system. Based on considerations on black hole
thermodynamics (bear in mind that the Bekenstein--Hawking entropy
is $S_{BH} = A/(4 \, \ell_{Pl}^{2})$, where $A$ is the area of the
black hole horizon), Bekenstein \cite{bekenstein} argued that the
maximum entropy for a box of volume $L^{3}$ should be proportional
to its surface rather than to its volume. In keeping with this, `t
Hooft conjectured that all phenomena within a volume $L^{3}$
should be described by a set of degrees of freedom located at the
surface which bounds this volume with approximately one binary
degree of freedom per Planck's area.

Inspired by these ideas, Cohen {\it et al.} \cite{cohen}
demonstrated that an effective field theory that saturates the
inequality (the Bekenstein bound)
\begin{equation}
L^{3}\, \Lambda^{3} \leq S_{BH} \simeq L^{2}M_{Pl}^{2}\,
\quad\qquad (M_{Pl}^{2} = (8\pi G)^{-1})\, ,
 \label{saturates}
\end{equation}
necessarily includes states for which the Schwarzschild radius
$R_{s}$ is larger than the box size, i.e., $R_{s} > L$. Namely,
for sufficiently high temperatures (in Planck units $T \gg L^{-1}$
but $T \leq \Lambda$) the thermal energy of the system is $E
\simeq L^{3}T^{4}$ while its entropy is $S\simeq L^{3}T^{3}$. When
Eq. (\ref{saturates}) is saturated (by setting $T = \Lambda$ in
there) it follows that $T \simeq(M_{Pl}^{2}/L)^{1/3}$. Now, the
Schwarzschild radius $R_{s}$ is related to the energy by $R_{s}
\simeq E/M_{Pl}^{2}$ (recall that an object with Schwarzschild
radius $R_{s}$ corresponds to a (Newtonian) mass $R_{s}/(2G)$).
Consequently, $R_{s} \gg L$, i.e., the Schwarzschild radius is
indeed larger than the system size.

A stronger constraint which excludes states for which the
Schwarzschild radius exceeds the size $L$ is
\begin{equation}
L^{3}\, \Lambda^{4} \leq M_{Pl}^{2}\, L \, .  \label{DEineq}
\end{equation}
The expression on the right-hand side of  Eq. (\ref{DEineq})
corresponds to the energy of a black hole of size $L$. So, this
constraint ensures that the energy $ L^{3}\Lambda^{4}$ in a box of
the size $L$ does not exceed the energy of a black hole of the
same size \cite{mli1}.

While the Bekenstein bound (\ref{saturates}) implies a scaling $L
\propto \Lambda^{-3}$, the inequality (\ref{DEineq}) corresponds
to a behavior $L \propto \Lambda^{-2}$. Furthermore, since
saturation of (\ref{DEineq}) means $\Lambda^{3} \simeq
(M_{Pl}^{2}/L^{2})^{3/4}$, one finds that $S = L^{3}\, \Lambda^{3}
= (M_{Pl}^{2}\, L^{2})^{3/4} = S_{BH}^{3/4}$.

By saturating the inequality (\ref{DEineq}) and identifying
$\Lambda^{4}$ with the holographic energy density $\rho_{X}$ it
follows \cite{cohen,mli1}
\begin{equation} \rho_{HDE}= \frac{3 c^{2}}{8\pi G \, L^{2}}\, ,
\label{rhox}
\end{equation}
where the factor $3$ was introduced for convenience and $c^{2}$ is
a dimensionless quantity which is usually  assumed constant. It is
obvious, that for a choice $L = H^{-1}$, i.e., for a cutoff set by
the Hubble radius, the energy density (\ref{rhox}) is
characterized  by exactly that dependence $\propto H^{2}$ which
was singled out by Eqs.~(\ref{rdotalt2}) and (\ref{rdotalt3}). In
the following we shall identify (\ref{rhox}) with the energy
density $\rho_{X}$ appearing in Eq.~(\ref{dotrhom}).

\section{Holographic dark energy in interaction with matter}
\label{interaction}

For holographic dark energy models with a cutoff scale $H^{-1}$
the time dependence of the energy density ratio is determined by
Eq.~(\ref{rdotalt3}). For a spatially flat universe the ratio $r$
remains constant. Possible changes of $r$ due to a non-vanishing
spatial curvature term are necessarily small. At first glance,
models with a constant (or almost constant) energy density ratio
seem  unable to account for a transition from decelerated to
accelerated expansion. And indeed, holographic dark energy models
with the Hubble scale as cutoff length were considered to be ruled
out since their equation of state parameter does not seem to allow
negative values which are required for an accelerated expansion
\cite{Hsu,mli1}. However, as shown previously, the presence of an
interaction between dark energy and dark matter  may change this
situation \cite{DW}. This becomes obvious if Eq.~(\ref{2dotH/H})
with $p = w \rho_{X}$ is solved for the equation of state
parameter $w$. The result is
\\
\begin{equation}
w  = - \frac{1}{1 - \Omega_{X}}\,\left(\frac{\Gamma}{3H} +
\frac{\Omega_{k}}{3} \right)\ .\label{w}
\end{equation}
\\
Assuming $|\Omega_{k}|$ small, the interaction rate $\Gamma$
essentially determines the equation of state parameter as soon as
the dimensionless ratio $\Gamma/H$, which here and in the
following is the relevant interaction parameter, becomes of order
one. For negligible interaction, $|\Gamma|/H \ll 1$, one has $|w|
\ll 1$.

The main point of interest now is the behavior of the deceleration
parameter, i.e., whether  this kind of models admits a transition
from decelerated to accelerated expansion. The condition for
accelerated expansion ($ q< 0$)
\begin{equation}
w + \frac{\Gamma}{3H}  < -%
\frac{1}{3} \, , \label{acchol}
\end{equation}
readily follows from the above equations. Introduction of the
expression (\ref{w}) for $w$ into the latter leads to
\begin{equation}
\frac{\Gamma}{H}\, %
\, >\, \frac{1 -\Omega_{X}}{\Omega_{X}}-
\frac{\Omega_{k}}{\Omega_{X}}\, \quad \Leftrightarrow \quad
\frac{\Gamma}{H} > r\ , \label{acch}
\end{equation}
where we have used Friedmann's equation (\ref{sumOmega}). On the
other hand, since holographic energy is not compatible with
phantom fields \cite{no-compatible}, one must impose $w \geq -1$
to be consistent with the holographic idea. From Eq. (\ref{w}) we
find that this condition amounts to
\begin{equation}
\frac{\Gamma}{H}  \leq 3\,\left(1 - \Omega_{X}\right) - \Omega_{k}
\, .\label{w>}
\end{equation}
Therefore, the consistency condition simply reads,
\begin{equation}
3 \Omega_{X}
> 1 - \Omega_{k} \, . \label{omegax>}
\end{equation}
The inequality (\ref{w>}) says that the ratio on its left hand
side is smaller than a quantity of  order one. Clearly, this is
compatible with an initial condition $|\Gamma|/(3H)\ll 1$. The
latter implies that $w$ must be close to zero deep in the matter
dominated era (cf. Eq.~(\ref{w})). An initial condition $|\Gamma|
\ll H$ is equivalent to an initially negligible interaction. This
means that at high redshifts (but well after matter--radiation
equality), the dark energy component is almost non--interacting
and its equation of state is close to that of pressureless matter.
On the other hand, the condition (\ref{acch}) for accelerated
expansion entails that $|\Gamma|/H$ must be larger than a lower
bound of order one, as well. Both inequalities are consistent with
condition (\ref{omegax>}). This supports our suggestion that,
depending on the interaction rate, a transition from decelerated
to accelerated expansion is indeed feasible.

Starting from Eq.~(\ref{rdotalt2}), the dynamics of the energy
density ratio may alternatively be written as
\begin{equation}
\dot{r} =  H\,\frac{\Omega_{k}}{1 - \Omega_{X}} \,\left[
\frac{\Gamma}{H}- r\right]\ .\label{rdothol1}
\end{equation}
For $k=0$ we recover the stationary case $r=$ constant,
irrespective of the value of $\Gamma/H $. If additionally
$\Gamma/H =$ constant, it follows from Eq. (\ref{w}) that $w$ is
also a constant. In such a case, there is no transition from
decelerated to accelerated expansion \cite{DW}. However, if
$\Gamma/H$ is allowed to grow, then the transition may well occur,
even though the ratio $r$ remains constant or almost constant (see
Eq.~(\ref{diffratio}) below).

The spatially curved cases generate an interesting additional
dynamics. The transition condition $\Gamma/H = r$ (cf. Eq.
(\ref{acch})) coincides with the condition for $\dot{r} = 0$ for
$k = \pm 1$ in (\ref{rdothol1}). Since at early times one has
$\Gamma/H \ll r$, the ratio $r$ in (\ref{rdothol1}) increases for
$k = +1$,  while it decreases for $k = - 1$. It reaches a maximum
(minimum) for $k = 1$ ($k = -1$) at $\Gamma/H = r$ and may go down
(up) afterward.

The obvious requirement for a sufficient growth of the interaction
parameter $\Gamma/H$ is that $(\Gamma/Hr)^{\displaystyle \cdot}%
>0$ in the phase of decelerated expansion  or, by (\ref{rdothol1}),
that
\\
\begin{equation}
\frac{\left(\Gamma/H\right)^{\displaystyle \cdot}}{\Gamma/H}> - \frac{H \, \Omega_{k}}{1 -%
\Omega_{X}}\,\left[1 -\frac{1}{r}\, \frac{\Gamma}{H}\right] \ .
 \label{dotGammar1}
\end{equation}
\\
But according to Eq. (\ref{w>}) the ratio $\Gamma/H $ is bounded
from above. For a growth of $\Gamma/H$ from a very small value to
a value of order one, a transition from decelerated to accelerated
expansion is feasible. We mention again, that any change of $r$
arises solely as a curvature effect.

A different way to understand the dynamics of the ratio $r$, which
also may be seen as a consistency check, is to start from
Eq.~(\ref{sumOmega}) and to realize that
$\Omega_{X}= $ constant. Then (recall that $r =%
\Omega_{M}/\Omega_{X}$),
\begin{equation}
r = \frac{1 - \Omega_{X}}{\Omega_{X}} -
\frac{\Omega_{k}}{\Omega_{X}} \, . \label{rconst+}
\end{equation}
The first term on the right hand side of (\ref{rconst+}) is
constant. It is just the curvature term that makes $r$ vary. Since
in accelerating universes $|\Omega_{k}|$ decreases $r$ will
approach a constant value in the long time limit. The ratio $r$
grows if the curvature term enlarges the right hand side ($k = 1$
and $q > 0$, or $k = -1$ and $q < 0$) and it goes down if the
curvature term diminishes the right hand side ($k = 1$ and $q <
0$, or $k = -1$ and $q > 0$). This is essentially the behavior
discussed beneath Eq.~(\ref{rdotalt4}). Clearly, differentiation
of Eq. (\ref{rconst+}) alongside the introduction of the
deceleration parameter $q$ consistently reproduces
Eq.~(\ref{rdotalt4}).

Combining Eqs.~(\ref{rdotalt4}) and (\ref{rdothol1}) we get
\begin{equation}
\frac{\Gamma}{H} - r = - 2\,\frac{1 - \Omega_{X}}{\Omega_{X}}\,q \
.\label{diffratio}
\end{equation}
This equation encodes the central message of our paper. It covers
all the cases $k= 0, \pm 1$ and explicitly shows that the
cosmological dynamics crucially depends on the difference of two
ratios: the ratio $\Gamma/H$  and the ratio $r$ of the energy
densities. The difference of these two ratios is directly
proportional to the (negative) deceleration
parameter. Accelerated expansion  requires $\Gamma/H%
> r$, decelerated expansion demands $\Gamma/H < r$.
It is obvious, that even a constant or a slowly varying ratio $r$
may be compatible with a sign change of $q$ provided $\Gamma/H$
evolves accordingly.  Even with a constant $r$ a present
accelerated expansion is compatible with an earlier matter
dominated period with decelerated expansion. The point is that for
negligible interaction (at sufficiently high redshifts) the dark
energy component behaves as non-relativistic matter. In this phase
$r$ is not really an important parameter since it describes the
ratio of two components with the same equation of state. It was
this property that apparently ruled out a (non-interacting)
holographic dark energy model with an infrared cutoff set by the
Hubble scale \cite{Hsu,mli1}. Here, this unwanted (in the
non-interacting model) feature is advantageous since, thanks to
it, a matter dominated phase during which structure formation can
occur is naturally recovered. It is only because of  the gradually
increased interaction that the equations of state begin to differ
from each other. Then $r$ becomes important.  At the first glance
the circumstance that $r$ is constant or only slowly varying seems
to imply, that the coincidence problem gets significantly
alleviated. But in fact, it has  been shifted from the problem to
explain a constant (or slowly varying) ratio $r$ towards the
problem to explain an interaction rate which has to be of the
order of the Hubble rate just at the present time. We argue that
an alternative, dynamical formulation of the coincidence problem
maybe useful, in particular, if a specific interacting model of
the type presented here will lead to potentially observable
differences from, say the $\Lambda$CDM model or generalized
Chaplygin gas models (see below).

One might argue against a constant -or nearly constant- ratio $r$
by saying that, as is well known, at the time of primeval
nucleosynthesis  the dark energy should not contribute more than
$5$ per cent to the total energy density  if the standard big bang
scenario for the build up of light elements is to hold
\cite{rachelbean}. However, this criticism does not apply to our
case since, as said above, the equation of state of dark energy,
$w$,  was not close to $-1$ at early epochs but close to that of
dust.

By starting from the expression (\ref{rhox}) with $L = H^{-1}$ for
the dark energy we have assumed the parameter $c^{2}$ to be
constant throughout our considerations. However, there does not
seem to exist any compelling reason for the inequality
(\ref{DEineq}) to be saturated or for the degree of saturation to
be a constant once and for all. In principle, a weak time
dependence ($0 < \left(c^{2}\right)^{\displaystyle \cdot}/c^{2}
\ll H$) of $c^{2}$ should be admitted here. Such an additional
degree of freedom will modify the dynamics discussed so far. In
particular, instead of described by Eq. (\ref{rdothol1}), the
ratio $r$ will evolve according to
\\
\begin{equation}
\dot{r} = - H\,r\,\frac{1}{1 - \Omega_{X}}\,\left\{- \Omega_{k}
\,\left[\frac{1}{r}\, \frac{\Gamma}{H}- 1\right] +
\frac{\left(c^{2}\right)^{\displaystyle \cdot}}{H\,c^{2}}
\right\}\, , \label{rdothol1c}
\end{equation}
\\
and the equation of state parameter changes from (\ref{w}) to
\\
\begin{equation}
w  = - \frac{1}{1 - \Omega_{X}}\,\left[\frac{\Gamma}{3H} +
\frac{\Omega_{k}}{3} + \frac{1}{3H}\,
\frac{\left(c^{2}\right)^{\displaystyle \cdot}}{c^{2}}\right]\
.\label{wc}
\end{equation}
\\
As a consequence, in the conditions (\ref{acchol}) and
(\ref{acch}) and in all relations after them, the rate $\Gamma$ is
to be replaced by  $\Gamma%
+\frac{\left(c^{2}\right)^{\displaystyle \cdot}}{c^{2}}$. A
varying $c^{2}$ will induce small variations in $r$, additionally
to those which are due to the spatial curvature and can support
the transition from decelerated to accelerated expansion \cite{DW}
(see also \cite{raul2}). Thus, at this transition, we have
\\
\begin{equation}
\dot{r}(q=0) = -\frac{r}{1 - \Omega_{X}}\,
\frac{\left(c^{2}\right)^{\displaystyle \cdot}}{c^{2}}\,\left(1 +
\frac{\Omega_{k}}{r}\right)\ ,\label{rdothol2c}
\end{equation}
\\
which reduces to the previous result $\dot{r}(q=0) =0$ for $c^{2}
= $ constant.

Before closing this section, it is noteworthy   that a negative
equation of state parameter $w$ can be obtained from a negative
curvature term ($k = -1$) alone (cf. Eq.~(\ref{w})), i.e., even
for $\Gamma = 0$,
\\
\begin{equation}
w  = - \frac{1}{3}\,\frac{1}{a^{2}H^{2}}\frac{1}{1 - \Omega_{X}}\,
. \label{wk-}
\end{equation}
\\
However, the condition (\ref{acch})  for accelerated expansion, in
this case $1/(a\,H)^{2} > 1 - \Omega_{X}$, can never be satisfied
as it contradicts the constraint equation (\ref{sumOmega}).
Consequently, only models with a non-vanishing interaction between
dark energy and dark matter can be potential candidates for a
satisfactory cosmological dynamics based on the holographic idea
with the infrared cutoff set by the Hubble function.

\section{A simple model}
\label{model} In our approach, the unknown nature of dark energy
is mapped on a (so far) unspecified interaction in the dark sector
quantified by the ratio $\Gamma / H$. In this section we build a
simple model that explicitly exhibits the general features
discussed in the preceding section.

With the help of the equation of state parameter (\ref{w}) the
balances expressions (\ref{dotrhom}) may be written as
\\
\begin{equation}
\frac{\dot{\rho}_{X}}{\rho_{X}} = - 3H \left[1 - \frac{1}{3}\,
\frac{\Omega_{k}}{1 - \Omega_{X}} - \frac{\Gamma}{3H}\,
\frac{\Omega_{X}}{1 - \Omega_{X}}\right] \, , \quad
\frac{\dot{\rho}_{M}}{\rho_{M}} = - 3H \left[1  -
\frac{\Gamma}{3H}\frac{\Omega_{X}}{\Omega_{M}}\right] \ .
 \label{dotfrx}
\end{equation}
%\\
%and
%\\
%\begin{equation}
%\frac{\dot{\rho}_{M}}{\rho_{M}} = - 3H \left[1  -
%\frac{\Gamma}{3H}\frac{\Omega_{X}}{\Omega_{M}}\right] \ .
%\label{dotfrm}
%\end{equation}
\\
For $k = 0$, we have $1 - \Omega_{X} = \Omega_{M}$ and both rates
coincide. Moreover, in this case $r$ stays constant. Integration
of (\ref{dotfrx}.2) under this condition from some initial period
(subscript ``$i$") onward, results in
\\
\begin{equation}
\frac{\rho_{M}}{\rho_{Mi}} = \left(\frac{a_{i}}{a}\right)^{3}\,
\exp{\left[\frac{1}{r}\int \frac{d\,a}{a}\frac{\Gamma}{H}\right]}
 \ .
\label{rm}
\end{equation}
\\
The exponential describes the deviation from an Einstein - de
Sitter universe.
 For the special case of a constant $\Gamma/H $, the expression (\ref{rm}) reduces to
\\
\begin{equation}
\frac{\rho_{M}}{\rho_{Mi}} = \left(\frac{a_{i}}{a}\right)^{3 -
\Gamma/(rH)}\,
 \ .
\label{rmconst}
\end{equation}
\\
Consequently, the Hubble parameter depends on the scale factor as
$H \propto a^{-\frac{1}{2}\left[3 - \Gamma/(rH)\right]}$. Hence,
\begin{equation}
a \propto t^{\frac{2}{3- \Gamma/(rH)}} \ .\label{a}
\end{equation}
The exponent on the right hand side of the last expression is
larger than unity for $\Gamma/(rH) > 1$ which reproduces the
previous condition (\ref{acch}) for accelerated expansion. The
limiting case $\Gamma/(rH) = 3$ corresponds to a constant energy
density, i.e., to an exponential growth of the scale factor.

As mentioned above, a constant ratio $\Gamma/ H$ can describe
either a phase of decelerated expansion ($\Gamma/(rH) < 1$), or a
phase of accelerated expansion ($\Gamma/(rH) > 1$), but never a
transition between the two. To stage a transition from decelerated
to accelerated expansion  $\Gamma/(rH)$ must increase which, for a
constant ratio $r$, means that $\Gamma/ H$ must increase. Since
there is no guidance of what a microscopic interaction model could
be, we shall resort to a phenomenological approach, assuming a
growth of the ratio $\Gamma/ H$ with a power of the scale factor
$a$. Notice that a working microscopic interaction model in the
present context, where it is exclusively the interaction that
drives the accelerated expansion, would be equivalent to
explaining the nature of dark energy. The fact that we obtain a
transition from decelerated to accelerated expansion under the
condition of a constant energy density ratio makes the coincidence
problem appear in a different way. In our approach it remains to
explain why the interaction rate is of the order of the Hubble
rate just at the present time. We argue, that this reformulation
of the coincidence problem may be advantageous since it offers a
potential dynamical solution, albeit hypothetical at the present
state of knowledge.

In the following we assume a growth according to
\\
\begin{equation}
\frac{\Gamma}{r H} = 3\beta \left(\frac{a}{a_{0}}\right)^{\alpha}
\ , \qquad a \leq a_{0} , \quad\beta\lesssim 1\, ,
\label{ansgamma}
\end{equation}
\\
with constant, positive--definite parameters $\alpha$ and $\beta$.
The condition on $\beta$ ensures, that the phantom divide cannot
be crossed. Furthermore, we restrict our considerations to the
past evolution of the universe. The ansatz (\ref{ansgamma})
implies a continuous growth of the ratio $\Gamma/(r H)$ with a
maximum value at the present time. We do not speculate about a
possible influence of the interaction on the future dynamics of
the Universe. But it is obvious that if, in the future, the
interaction becomes less effective, an evolution back to a fresh
phase of decelerated expansion is feasible.

Inserting the ansatz (\ref{ansgamma}) into Eq. (\ref{rm}) we get
\\
\begin{equation}
\frac{\rho_{M}}{\rho_{Mi}} = \left(\frac{a_{i}}{a}\right)^{3}\,
\exp{\left[\frac{3\beta}{\alpha}
\left(\left(\frac{a}{a_{0}}\right)^{\alpha} -
\left(\frac{a_{i}}{a_{0}}\right)^{\alpha}\right)\right]}
 \ .
\label{rmans}
\end{equation}
\\
The exponential factor diminishes the decreases of the energy
density with cosmic expansion. Replacing the initial quantities by
the corresponding present day values (subscript ``0") provides us
with
\\
\begin{equation}
\rho_{M} = \rho_{M0}\left(\frac{a_{0}}{a}\right)^{3}\,
\exp{\left[\frac{3\beta}{\alpha}
\left(\left(\frac{a}{a_{0}}\right)^{\alpha} -
 1\right)\right]}
 \ .
\label{rmind0}
\end{equation}
\\
For $k=0$ and $c^{2} = $ constant, the energy density of the dark
energy component shows exactly the same dependence on the scale
factor,
\\
\begin{equation}
\rho_{X} = \rho_{X0}\left(\frac{a_{0}}{a}\right)^{3}\,
\exp{\left[\frac{3\beta}{\alpha}
\left(\left(\frac{a}{a_{0}}\right)^{\alpha} -
 1\right)\right]} \, .
\label{rxind0}
\end{equation}
\\
Thus, the Hubble rate is given by
\\
\begin{equation}
h \equiv \frac{H}{H_{0}}= \left(\frac{a_{0}}{a}\right)^{3/2}\,
\exp{\left[\frac{3\beta}{2\alpha}
\left(\left(\frac{a}{a_{0}}\right)^{\alpha} -
 1\right)\right]}
 \ .
\label{H/H0}
\end{equation}
\\
Comparing Eq. (\ref{ansgamma}) with (\ref{diffratio}) for the
spatially flat case, the deceleration parameter is found to be
\\
\begin{equation}
q = \frac{1}{2} - \frac{3}{2}\,
\beta\left(\frac{a}{a_{0}}\right)^{\alpha}
 \ .
\label{qbeta}
\end{equation}
\\
It is expedient to compare the dimensionless Hubble rate, $h$,
given by Eq. (\ref{H/H0}), with the corresponding quantity of the
$\Lambda$CDM model,
\\
\begin{equation}
\frac{H_{\Lambda CDM}}{H_{0}} =
\sqrt{\frac{\rho_{\Lambda}}{\rho_{\Lambda} + \rho_{M0}}}\left[1 +
\frac{\rho_{M0}}{\rho_{\Lambda}}\left(\frac{a_{0}}{a}\right)^{3}\right]^{1/2}
 \ .
\label{H/H0LCDM}
\end{equation}
\\
Introducing the redshift parameter $z$ by $a_{0}/a = 1 + z$, we
have
\\
\begin{eqnarray}
\frac{H_{\Lambda CDM}}{H_{0}} &=&
\sqrt{\frac{\rho_{\Lambda}}{\rho_{\Lambda} + \rho_{M0}}}\left[1 +
\frac{\rho_{M0}}{\rho_{\Lambda}}\left(1 +
z\right)^{3}\right]^{1/2}\nonumber\\ &=& 1 +
\frac{3}{2}\frac{\rho_{M0}}{\rho_{\Lambda} + \rho_{M0}}\, z +
{\cal O}(z^{2}) \, ,
\label{H/H0LCDMz}
\end{eqnarray}
\\
where for  reasons  explained below we have retained only the
linear term in $z$ in the second line of last equation. From
(\ref{H/H0}) it follows that
\\
\begin{figure}[th]
\includegraphics[width=3.5in,angle=-90,clip=true]{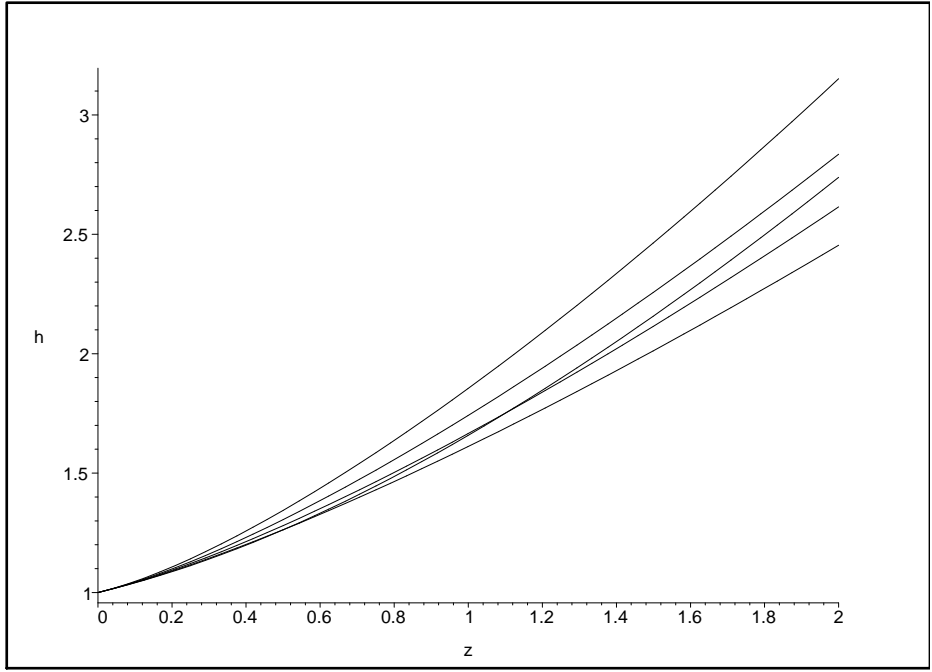}
\caption{Graphs of the dimensionless ratio $h$ vs redshift. Except
for the middle graph -which corresponds to the $\Lambda$CDM
model-, from top to bottom the $\alpha$ values are: $2.0$, $1.5$,
$1.2$, and $1.0$. Here, we have taken $\beta= 3/4$.}
\label{fig:toy1}
\end{figure}
\\
\begin{eqnarray}
\frac{H}{H_{0}} &=& \left(1 + z\right)^{3/2}\,
\exp{\left[\frac{3\beta}{2\alpha} \left(\left(\frac{1}{1 +
z}\right)^{\alpha} -
 1\right)\right]}\nonumber\\
 &=& 1 + \frac{3}{2}\left(1 - \beta\right)z + {\cal O}(z^{2}) \, ,
\label{H/H0z}
\end{eqnarray}
for our interacting model. It is remarkable that to linear order
in $z$, the expression (\ref{H/H0z}) does not depend on $\alpha$.

We may now require that our model (\ref{H/H0z}) coincides with the
$\Lambda$CDM model (\ref{H/H0LCDMz}) up to linear order in $z$.
This allows us to fix the parameter $\beta$ in terms of the
observationally well established ratio of the parameters
$\rho_{M0}$ and $\rho_{\Lambda}$ of the $\Lambda$CDM model. The
result is
\begin{equation}
\beta = \frac{\rho_{\Lambda}}{\rho_{\Lambda} + \rho_{M0}}
 \ .
\label{beta}
\end{equation}
For the observed $\rho_{M0}/\rho_{\Lambda} \approx 1/3$ we have
$\beta \approx 3/4$. Under this condition, the present value of
the deceleration
parameter (\ref{qbeta}), $\; q_{0} = \frac{1}{2} - %
\frac{3}{2}\beta$, coincides with the corresponding value
$q_{\Lambda CDM 0} = 1/2%
-[3\rho_{\Lambda}/2(\rho_{\Lambda}+\rho_{M0})] \approx - 0.6$ of
the $\Lambda$CDM model as well. However, the evolution of $q$
towards this value depends on $\alpha$ which gives rise to
differences in any order beyond the linear one. For the value of
the transition redshift $z_{acc}$, that follows from $q=0$ in
(\ref{qbeta}), we obtain $z_{acc} = \left(3\beta\right)^{1/\alpha}%
- 1$.

With the identification (\ref{beta}) and $\rho_{M0}/\rho_{\Lambda}
\approx 1/3$ we find $z_{acc} \approx 1.2$ for $\alpha = 1$ and
$z_{acc} \approx 0.5$ for $\alpha = 2$. The corresponding
$z$-value for the $\Lambda$CDM model is approximately $0.8$.

Figure 1 contrasts the Hubble parameter prediction of our
interaction model for different values of $\alpha$ with the
$\Lambda$CDM model. Clearly, the free parameter $\alpha$,
obviously of the order of one, can be used to adjust $h(z)$
according to the observational situation.

Figure 2 presents the best fit of our model and the best fit of
the $\Lambda$CDM model to the Gold SNIa data set of Riess {\em et
al.} (fourth reference in \cite{SNIa}), and the SNLS data set of
Astier {\em et al.} (sixth reference in \cite{SNIa}). In plotting
the graphs the distance modulus, $\mu = 5\, \log d_{L} +25$,  was
employed. In this expression $d_{L} = (1+z)
\int_{0}^{z}{H^{-1}(z')\, dz'}$ is the luminosity distance in
megaparsecs. As can be seen, both best fits largely overlap one
another. In all, current SNIa data are unable to discriminate
between the popular $\Lambda$CDM and our interaction model.
 \\
\begin{figure}[th]
\includegraphics[width=4.5in,angle=0,clip=true]{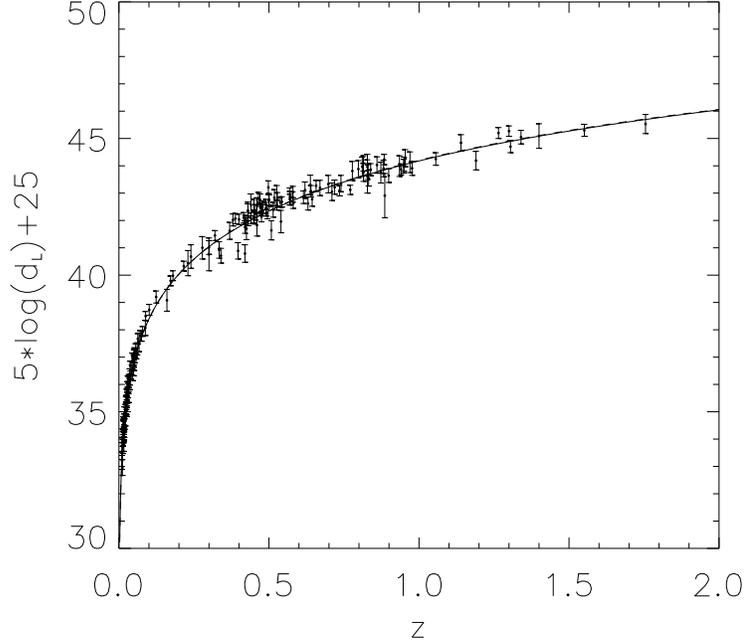}
\caption{Distance modulus vs redshift for the best fit model,
$\Omega_{X} = 0.75$, $H_{0} = 67.35$ km/s/Mpc, $\alpha = 1.5$
(solid line), and the $\Lambda$CDM model, $\Omega_{\Lambda} =
0.71$, and $H_{0} = 67.35$ km/s/Mpc (dashed line).}
\end{figure}

\section{Perturbation dynamics}
\label{perturbations}

In a flat universe the dark energy equation of state parameter is
given by $w = -\left(1 + r\right)\frac{\Gamma}{3Hr}$. Since
$\rho_{X} = \frac{\rho}{1 + r}$, the total equation of state of
the cosmic medium is
\\
\begin{equation}
\frac{p}{\rho} = - \frac{\Gamma}{3Hr}
 \ .
\label{toteos}
\end{equation}
\\
The corresponding adiabatic sound speed parameter becomes
\\
\begin{equation}
\frac{\dot{p}}{\dot{\rho}} = \left(1 -\frac{1}{3 \gamma
H}\frac{\left(\Gamma/H\right)^{\displaystyle\cdot}}{\Gamma/H}\right)\frac{p}{\rho}\
, \qquad\qquad \qquad\left(\gamma = 1 + \frac{p}{\rho}\right)
 \ .
\label{adsound}
\end{equation}
\\
However, the interaction introduces non-adiabatic features into
the perturbation dynamics. Deviations from the adiabatic behavior
are suitably characterized by the quantity $\hat{p} -
\frac{\dot{p}}{\dot{\rho}}\hat{\rho}$, which in our case takes the
form
\\
\begin{equation}
\hat{p} - \frac{\dot{p}}{\dot{\rho}}\hat{\rho} = p
\left[\frac{\left(\Gamma/H\right)^{\hat{}}}{\Gamma/H} -
\frac{\left(\Gamma/H\right)^{\displaystyle\cdot}}{\Gamma/H}
\frac{\hat{\rho}}{\dot{\rho}} - \frac{\hat{r}}{r}\right]
 \ ,
\label{nad}
\end{equation}
\\
where a hat indicates perturbation of the corresponding quantity.
The contribution on the right-hand side of (\ref{nad}) accounts
both for the non-adiabaticity of the $X$ component and for a
contribution which results from the two component nature of the
system. In terms of the components this expression is equivalent
to
\\
\begin{equation}
P - \frac{\dot{p}}{\dot{\rho}} D =
\frac{\dot{\rho}_{X}}{\dot{\rho}}\left(P_{X} -
\frac{\dot{p}_{X}}{\dot{\rho}_{X}} D_{X}\right) +
\frac{\dot{\rho}_{M}\dot{\rho}_{X}}{\dot{\rho}^{2}}\,
\frac{\dot{p}_{X}}{\dot{\rho}_{X}}\left[D_{X} - D_{M}\right]
 \ ,
\label{nadPD}
\end{equation}
\\
where
\begin{equation}
D \equiv  \frac{\hat{\rho}}{\rho+ p} = - 3H
\frac{\hat{\rho}}{\dot\rho}\ , \label{D}
\end{equation}
and  the energy density perturbations for the components are
\begin{equation}
D_{M} \equiv  - 3H \frac{\hat{\rho}_M}{\dot\rho_M}\, \quad\mathrm{and} \quad D_{X}
\equiv  - 3H \frac{\hat{\rho}_X}{\dot\rho_X}
 \
.\label{DA}
\end{equation}
Analogously, the pressure perturbations are described by
\begin{equation}
P \equiv  \frac{\hat{p}}{\rho+ p} = - 3H \frac{\hat{p}}{\dot\rho}
\, \quad \mathrm{and} \quad P_{X} \equiv  - 3H
\frac{\hat{p}_X}{\dot\rho_X} \label{Ptot}\ .
\end{equation}
The circumstance that $\dot{r} = 0$ in these class of models
results in a particularly simple structure for $D_{X} - D_{M}$.
Just from the definition of these quantities we find by direct
calculation
\begin{equation}
D_{X} - D_{M} = - \frac{\hat{r}}{\gamma r}
 \ .
\label{DX-DM}
\end{equation}
This difference is directly related to the fluctuation of the
energy density ratio. The dynamic equation for $D_{X} - D_{M}$
(see below) then determines the time dependence of $\hat{r}$.

A homogenous and isotropic background universe with scalar metric
perturbations and vanishing anisotropic pressure can be
characterized by the line element (longitudinal gauge, cf.
\cite{LL})
\begin{equation}
\mbox{d}s^{2} = - \left(1 + 2 \psi\right)\mbox{d}t^2 +
a^2\,\left(1-2\psi\right)\delta _{\alpha \beta}
\mbox{d}x^\alpha\mbox{d}x^\beta \ .\label{metric}
\end{equation}
The perturbation dynamics is most conveniently described in terms
of the gauge-invariant variable \cite{Bardeenetal83}
\begin{equation}
\zeta \equiv -\psi + \frac{1}{3}\frac{\hat{\rho}}{\rho + p} =
-\psi - H \frac{\hat{\rho}}{\dot\rho}\ ,
 \label{defzeta}
\end{equation}
which represents curvature perturbations on hypersurfaces of
constant energy density. Corresponding quantities for the
components are
\begin{equation}
\zeta_A \equiv - \psi - H \frac{\hat{\rho}_A}{\dot\rho _A} \quad
\qquad (A = X, M).
\label{defzetaA}
\end{equation}
On large perturbation scales the variable $\zeta$ obeys the
equation (cf. \cite{Lyth,GarciaWands,Wandsetal00})
\begin{equation}
\dot\zeta = - H \left(P - \frac{\dot{p}}{\dot{\rho}}D\right)  \ .
 \label{dotzetageneral}
\end{equation}
The equation for $\zeta_X - \zeta _M = \frac{1}{3} \left(D_{X}
-D_{M}\right)$ is
\begin{eqnarray}
\left(\zeta_X - \zeta _M \right)^{\displaystyle\cdot} &=&  3H^2
\frac{\hat{p}_X -
\frac{\dot{p}_X}{\dot{\rho}_X}\hat{\rho}_X}{\dot{\rho}_X} \nonumber\\
&& +  3H \frac{\dot\Pi}{\dot\rho_M} \frac{\dot\rho}{\dot\rho _X}
\left[ \zeta_\pi - \zeta + \frac{\dot\rho _X -
\dot\rho_M}{\dot\rho} \left(\zeta_X - \zeta_M\right)
\right] \ .
\label{dotzeta-}
\end{eqnarray}
The gauge invariant quantity $\zeta_\Pi$ describes the
perturbation of the interaction term. It is defined in analogy to
the other perturbation quantities,
\\
\begin{equation}
\zeta_{\Pi} \equiv - \psi - H \frac{\hat{\Pi}}{\dot\Pi}\ , \quad
\Pi = - \frac{\Gamma \rho_{X}}{3 H} \ .
 \label{zetapi}
\end{equation}
\\
Inserting in (\ref{dotzeta-}) the expression (\ref{DX-DM}) and
\\
\begin{equation}
\hat{p}_{X} - \frac{\dot{p}_{X}}{\dot{\rho}_{X}}\hat{\rho}_{X} = -
\frac{\Gamma\rho}{3 H r}
\left[\frac{\left(\Gamma/H\right)^{\hat{}}}{\Gamma/H} -
\frac{\hat{r}}{r \left(1 + r\right)} + \frac{1}{3\gamma H}
\frac{\left(\Gamma/H\right)^{\displaystyle\cdot}}{\Gamma/H}
\left(\frac{\hat{\rho}}{\rho} - \frac{\hat{r}}{1 + r}\right)
\right]
 \ ,
\label{nadX}
\end{equation}
\\
which follows from the equation of state (\ref{toteos}),
we obtain
\begin{equation}
\dot{\hat{r}} = 0 \quad \Rightarrow \quad \hat{r} = \mathrm{const}
 \
\label{hatrconst}
\end{equation}
on large perturbation scales, i.e. scales for which spatial
gradient terms may be neglected. This means, also at the
perturbative level, $\hat{r}$ is not a dynamical degree of freedom
on these scales. It is remarkable that this property holds for any
interaction rate. While, of course, the trivial case $\hat{r} = 0$
is included here, it is expedient to notice that any non-vanishing
$\hat{r}$, although constant, gives rise to non-adiabatic effects.

The simplest case to discuss this dynamics more specifically is
$\Gamma =$ const. Since $H \propto \rho^{1/2}$, this corresponds
to an effective background equation of state $p \propto -
\rho^{1/2}$ which is characteristic for a special generalized
Chaplygin gas \cite{Bento}. Its energy density is
\begin{equation}
\rho = \rho_{i}\left[\frac{\Gamma}{3 H_{i}r} + \left(1 -
\frac{\Gamma}{3
H_{i}r}\right)\left(\frac{a_{i}}{a}\right)^{3/2}\right]^{2}
 \ .
\label{rhoGCG}
\end{equation}
Again, for $\Gamma = 0$ the Einstein de Sitter universe is
reproduced. The general relations (\ref{adsound}) and (\ref{nad})
specify to
\begin{equation}
\frac{\dot{p}}{\dot{\rho}} = \frac{1}{2}\frac{p}{\rho}
 \
\label{adsound2}
\end{equation}
and
\begin{equation}
\hat{p} - \frac{\dot{p}}{\dot{\rho}}\hat{\rho} = \frac{1}{2} p
\left[\frac{\hat{\rho}}{\rho} - 2 \frac{\hat{\Theta}}{\Theta} -
\frac{\hat{r}}{r}\right] \quad \qquad (\Theta = 3 H)
 \ ,
\label{nad2}
\end{equation}
respectively. Only if spatial gradient terms in the perturbation
dynamics are neglected, the first two terms in the bracket on the
right-hand side of Eq.~(\ref{nad2}) cancel each other
approximately. Therefore one may expect that non-adiabatic effects
will be most important on small perturbation scales \cite{essay,bvm}. But even for
$\frac{\hat{\rho}}{\rho} \approx 2 \frac{\hat{\Theta}}{\Theta}$ a
non-vanishing perturbation $\hat{r}$ generates a non-adiabatic
dynamics of the system. The corresponding large scale perturbation
dynamics can be solved analytically. The equation for $\zeta$ in
this case reduces to
\begin{equation}
\dot{\zeta} = - \frac{\Gamma}{6 \gamma}\frac{\hat{r}}{r^{2}}
 \ .
\label{dotzr}
\end{equation}
With $\gamma = 1 + \frac{p}{\rho} = 1 - \frac{\Gamma}{3 Hr}$ we
may write
\begin{equation}
\zeta^{\prime} = -
\frac{\Gamma}{2}\frac{\hat{r}}{r}\frac{1}{a}\,\frac{1}{3 H r -
\Gamma}
 \ ,
\label{zetaprime}
\end{equation}
where $\zeta^{\prime} \equiv \frac{d\zeta}{d a}$. Since
(cf.~(\ref{rhoGCG}))
\begin{equation}
H = H_{i}\left[\frac{\Gamma}{3 H_{i}r} + \left(1 - \frac{\Gamma}{3
H_{i}r}\right)\left(\frac{a_{i}}{a}\right)^{3/2}\right]
 \ ,
\label{HGCG}
\end{equation}
equation (\ref{zetaprime}) is readily solved. The result is
\begin{equation}
\zeta = \zeta_{i} - \frac{\Gamma}{3}\frac{\hat{r}}{r}\frac{1}{3
H_{i}r - \Gamma}\left[\left(\frac{a}{a_{i}}\right)^{3/2} -
1\right]
 \ .
\label{zeta}
\end{equation}
The non-adiabatic contribution grows with $a^{3/2}$. This will
influence the time dependence of the gravitational potential and
hence the integrated Sachs-Wolfe effect (cf \cite{zimdahl,essay05}
for  similar studies).  Recall that the $\Lambda$CDM model is
characterized by a constant value of $\zeta$. On small scales we
expect a modification of the adiabatic sound speed
\cite{essay,bvm}. Notice that the quantity
$\frac{\dot{p}}{\dot{\rho}}$ in (\ref{adsound2}) is negative. As a
consequence, there occur small scale instabilities within an
adiabatic perturbation analysis of this generalized Chaplygin gas.
This apparently unrealistic feature has been used as an argument
to discard this type of models altogether \cite{tegmarkbis}. On
the other hand, it has been suggested that non-adiabatic effects
might modify this picture \cite{NJP,ioav,zimdahl}. That this is
indeed the case, has been demonstrated recently by an explicit
numerical analysis for the galaxy power spectrum \cite{bvm}. The
present consideration in the context of interacting holographic
models offer a systematic procedure to include such effects. We
plan to perform a more quantitative non-adiabatic perturbation
analysis in a subsequent work.

\section{Discussion}
\label{discussion} We have established a holographic dark energy
model in which a negative equation of state parameter as well as
the transition from decelerated to accelerated expansion arise
from pure interaction phenomena. A remarkable property of the
model is the fact that, in a spatially flat universe, the ratio of
the energy densities of dark matter and dark energy remains
constant during the transition. The non-interacting limit of this
model is an Einstein-de Sitter universe which is supposed to
describe our cosmos at high redshifts.  At first sight, the fact
that $r$ was never large may seem at variance with the
conventional scenario of cosmic structure formation as one may
think that at early times the amount of dark matter may have been
insufficient to  produce gravitational potential wells deep enough
to lead to the condensation of the galaxies. However, this is not
so; an earlier matter dominated phase is naturally recovered since
for negligible interaction at high redshifts the equation of state
of the dark energy is similar to that of non-relativistic matter.
In the context of our approach the coincidence problem is
dynamized and takes the form: ``why is the interaction rate that
drives the accelerated expansion of the order of the Hubble rate
precisely at the present epoch?" Based on the assumption that the
relevant interaction parameter grows with a power law in the scale
factor (Eq.~(\ref{ansgamma})), we have worked out a specific model
for the transition from decelerated to accelerated expansion under
the condition of a constant ratio of the energy densities of dark
matter and dark energy. A preliminary analysis shows that this
model fits the SNIa data not less well than the $\Lambda$CDM
model.

We have also shown that taking into account the spatial curvature
term gives rise to an additional dynamics which implies a small
(compared with the Hubble rate) change of the energy density
ratio.

Furthermore, we discussed to what extent a slowly varying
saturation parameter of the holographic bound may modify
the cosmological dynamics. Besides, our approach can cope with a
later transition to a new decelerated phase of expansion
\cite{carvalho}-something incompatible with holographic models
whose infrared cutoff is set by the radius of the future event
horizon.

The interaction between dark matter and dark energy introduces
non-adiabatic features into the perturbation theory. For the
special case of a constant decay rate of the dark energy, in the
background equivalent to a generalized Chaplygin gas, we find, Eq.
(\ref{zeta}), that the large scale curvature perturbations on
hypersurfaces of constant density (which are constant in the
adiabatic case, in particular for the $\Lambda$CDM model) vary
with the power $3/2$ of the scale factor.

\acknowledgments{W.Z. was supported by a Grant from the ``Programa
de Movilidad de Profesores Visitantes 2006" sponsored by the
``Direcci\'{o} General de Recerca de Catalunya". He also
acknowledges support by the Brazilian grants 308837/2005-3 (CNPq)
and 093/2007 (CNPq and FAPES). This work was partially supported
by the Spanish Ministry of Education and Science under Grant FIS
2006-12296-C02-01, the ``Direcci\'{o} General de Recerca de
Catalunya" under Grant 2005 SGR 000 87.}

\end{document}